# A Family of Constitutive Models Implemented in PLAXIS to Simulate Cemented Mine Backfill

Mauro SOTTILE[a,b], Nicolás LABANDA[a,b,1], Iñaki GARCÍA MENDIVE[a],
Osvaldo LEDESMA[a] and Alejo O. SFRISO[a,b]
[a] *SRK Consulting.*
[b] *University of Buenos Aires.*

**Abstract.** A family of constitutive models for mine cemented backfill is presented. Four formulas for the density- and pressure-dependency of elastic moduli, five formulas for the density- and pressure-dependency of friction angle and four formulas for the age-dependency of the elastic moduli and effective cohesion are incorporated into an isotropic hypoelasticity with Mohr-Coulomb perfect plasticity framework and implemented in PLAXIS as a user-defined material model. This family includes the standard Mohr-Coulomb, Bolton, Leps, Barton and Hoek-Brown models as trivial cases when both nonlinear elasticity and age-dependency are switched off. In this paper, the formulation of the models is introduced, the basis of the numerical implementation is outlined, and a case history of the application to the cemented backfill of a sublevel stoping mine is presented as an example.

**Keywords.** Constitutive modelling, cemented backfill, underground mining.

## 1. Introduction

Cemented rockfill and cemented paste have been widely used in the mining industry for backfilling underground openings to provide wall stability while mining adjacent stopes. Modelling mining processes incorporating cemented backfill requires a reliable estimate of the changes in the strength and stiffness of the material with   basis of the numerical implementation is outlined, and a case history of the application to the cemented backfill of a sublevel stoping mine is presented as an example. For brevity, formulas are introduced with little explanation; the derivations and support behind several of these formulas can be found in the relevant references.

## 2. Model formulation

### 2.1. State variables

State variables are age $t$ and void ratio $e$. Their update formulas are

$$t^{n+1} = t^n + \Delta t \tag{1a}$$

$$e^{n+1} = e^n + (1 + e^n)\Delta\epsilon_v \tag{1b}$$

where $t^n$ and $e^n$ are the age and void ratio at the beginning of the step, $\Delta t$ is the time increment and $\Delta\epsilon_v$ is the increment of volumetric strain.

---

[1] Nicolás A. Labanda, SRK Consulting Argentina, Chile 300, Buenos Aires, Argentina; E-mail: nlabanda@fi.uba.ar, nlabanda@srk.com.ar.

## 2.2. Elasticity

Standard isotropic hypoelasticity is employed. The elastic operator D is of the form

$$D = 2G\left(\frac{1+v}{1-2v} 1\otimes 1 + I^d\right) \tag{2}$$

where $G$ is the shear modulus, $v$ is Poisson's ratio, 1 is the 2nd order unit tensor, $\otimes$ the tensor product operator, and $I^d$ is the 4th order deviator projector tensor. Four expressions for $G$ were implemented:

$$G = G_{ref}\left((c_p - p)/p_{ref}\right)^m \quad [1] \tag{3a}$$

$$G = G_{ref}(c_e - e)^2/(1 + e)\left((c_p - p)/p_{ref}\right)^m \quad [2] \tag{3b}$$

$$G = G_{ref}/(c_e + (1 - c_e) e^2)\left((c_p - p)/p_{ref}\right)^m \quad [3] \tag{3c}$$

$$G = G_{ref}/e^{c_e}\left((c_p - p)/p_{ref}\right)^m \quad [4] \tag{3d}$$

where $G_{ref}[t]$ and $c_p[t]$ are age-dependent parameters, $c_e$ and $m$ are parameters, $p$ is mean pressure (tension positive), $e$ is void ratio and $p_{ref}$ is a reference pressure. Shear modulus in tension is equal to $G|_{p=0}$. The reader is referred to Ref. [1-4] for further discussion around these formulas.

## 2.3. Yield function and plastic potential

A standard Mohr-Coulomb yield surface with tension cut-off is employed. A curved envelope is obtained by employing density- and pressure-dependent formulas for the friction angle $\phi$, following the procedure described in [5]. Five expressions were implemented, namely constant friction angle, Leps/Barton [6, 7], Bolton [8], Sfriso [9], and Hoek-Brown [10].

$$\phi = \phi_0 \tag{4a}$$

$$\phi = \phi_0 - \Delta\phi \, log_{10}[\sigma_n] \geq \phi_{cv} \quad [6,7] \tag{4b}$$

$$\phi = \phi_{cv} + \Delta\phi\left(D_r\left(Q - ln[-100\, p/p_{ref}]\right) - 1\right) \leq \phi_{max} \leq \phi_{max} \quad [8] \tag{4c}$$

$$\phi = \phi_{cv} + \Delta\phi \left(D_r \; ln\left[(-p \; e^{2.5})/(p_r \; p_{ref})\right] - 1\right) \leq \phi_{max} \quad [9] \quad (4d)$$

$$\phi = sin^{-1}\left[3 \; m \; \sqrt{\sigma_c} / \left(\sqrt{m^2 \; \sigma_c + 36 \; s \; \sigma_c - 36 \; m \; p} + 2 \; m\sqrt{\sigma_c}\right)\right] \geq \\ \geq \phi_{cv} \quad [5] \quad (4e)$$

where $\phi_0$, $\Delta\phi$, $Q$, $p_r$, $m$, $s$ are parameters, $\sigma_c[t]$ is an age-dependent unconfined compression strength, $\phi_{cv}$ is the constant volume friction angle, $\phi_{max}$ is an upper limit of the friction angle, $D_r$ is relative density and $\sigma_n$ is the normal pressure in the sliding plane, computed in turn as a function of $p$. Age-dependent cohesion $c[t]$ is employed in the standard Mohr-Coulomb model (Eq. (4a)). Models derived from Eqs. (4b) to (4e) employ $\sigma_c[t]$ instead. The reader is referred to [5] for the derivation of Eq. (4) and details on the implementation of Hoek-Brown model as a Mohr-Coulomb-type model.

A standard Vermeer-deBorst plastic potential is employed. Except for the standard Mohr-Coulomb model which employs a constant value, the dilatancy angle is made dependent on confining pressure and density through the expression [8].

$$\psi = 0.8 \; (\phi - \phi_{cv}) < \psi_{max} \quad (5)$$

where $\psi_{max}$ is a limiting value. Strain-softening can occur if Eqs. (4c) or (4d) are employed in combination with Eq. (5). While strain-softening is realistic and does occur in dilating cemented materials, it produces a range of numerical issues including poor convergence and mesh-dependency in standard FEM models.

### 2.4. Age-dependency

Hardening of cement paste produces an increment in strength and stiffness that can be modelled by incorporating age-dependency in the input parameters (see for instance [11-14]). In this family of models, four evolution laws were implemented.

$$x = x_i \quad (6a)$$

$$x = x_i + (x_f - x_i) \; (t - t_i)/(t_f - t_i) \quad (6b)$$

$$x = (x_i - x_f) \; exp\left[-3 \; (t - t_i)/(t_f - t_i)\right] + x_f \quad (6c)$$

$$x = x_i + \frac{x_f - x_i}{2}\left(1 + tanh\left[log[19]\frac{t - (t_f + t_i)/2}{t_f - t_i}\right]\right) \quad (6d)$$

where $x$ stands for the relevant age-dependent parameter, and $x_i \mid x_f$ are the initial and final values at ages $t_i \mid t_f$. The user is able to switch on/off any combination of $G_{ref}$, $c_p$, $c$, and/or $\sigma_c$ as age-dependent parameters.

## 3. Numerical implementation

### 3.1. Integration in time

The mean value $\bar{x}$ of the relevant variable within the time step $\{t, t + \Delta t\}$ is computed by the exact (analytical) integration of the relevant Eq. (6a) to (6d).

$$\bar{x} = \int_{t}^{t+\Delta t} x[t]\, dt / \Delta t \qquad (7)$$

This integration is independent of the integration in strain and is performed before the latter, producing minimal impact in the efficiency of the numerical algorithm.

### 3.2. Integration in strain

A conventional, fully implicit integration scheme in strain is employed. The trial stress $\hat{\sigma}$ is computed assuming (nonlinear) elastic behavior by implicit integration of the relevant Eqs. (3a) to (3d).

Once $\hat{\sigma}$ is known, the elastic operator D (Eq. (2)), shear modulus (Eq. (3)), friction angle $\phi$ (Eq. (4)), dilatancy angle $\psi$ (Eq. (5)) and the yield function at the trial stress $f[\hat{\sigma}]$ are computed. If $f[\hat{\sigma}] \leq 0$, the strain step is (nonlinear) elastic; the updated stress is readily computed as $\sigma_{n+1} = \hat{\sigma}$ and reported back to the main program.

If $f[\hat{\sigma}] > 0$ the strain step is elastoplastic. In this case, the updated stress $\sigma_{n+1}$ is initialized $\sigma_{n+1}^{(0)} = \hat{\sigma}$, where (0) stands for the starting value of the iterator. A nested update algorithm is then employed as outlined in Figure 1. First, the functions of state variables are frozen at the given state. Then, a Mohr-Coulomb yield surface with perfect plasticity is employed for the stress update [15]. Once $\sigma^{n+1}$ is updated, the relevant functions of state variables are updated, and the loop is repeated until convergence, which is usually achieved in two to five iterations. The reader is referred to Ref. [5] for further details of this numerical implementation.

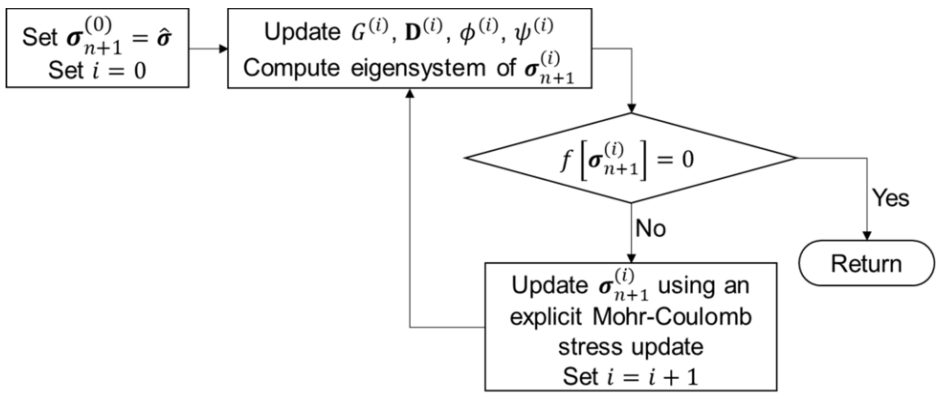

**Figure 1.** Flowchart of the algorithm for the stress update.

## 4. Application to a sublevel stoping mining case

*4.1. Description of the problem*

At an underground gold mine in South America the tabular orebody lies approximately 250 m below ground surface. The ore is recovered via sublevel stoping, stopes being 15 m long and 20 m high, and progressively back-filled with cemented rockfill (CRF).

A 2D plane strain numerical model was set up in Plaxis to evaluate the stress distribution due to the extraction sequence and to assess the adequacy of the CRF strength and stiffness to control deformations of the country rock.

*4.2. Geometry and mesh*

The model is 500 m wide and 170 m deep, has 6552 15-noded triangular elements with an average size of 3.34 m (Figure 2). The weight of the overlying 200 m of rock is modelled by a layer of equivalent weight.

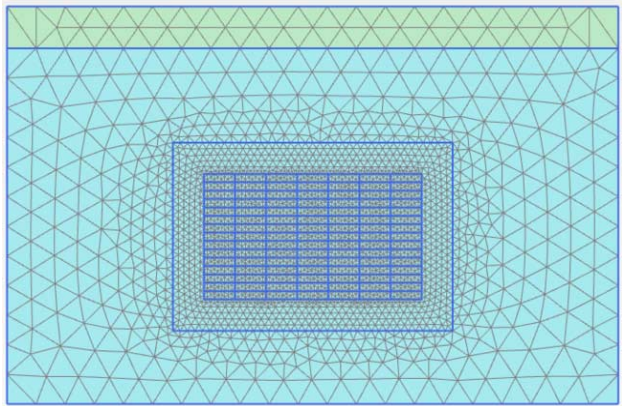

**Figure 2.** 2D finite element mesh.

*4.3. Materials*

The orebody was modelled using the Hoek-Brown model available in Plaxis. Both the cemented and uncemented backfills were modelled using Eq. (3a), Eq. (4a) and Eq. (6c), simplest choice for CRF. Age-independent parameters are presented in Table 1. See [16] for the description of the parameters of the Hoek-Brown criterion in Plaxis.

**Table 1.** Age-independent material parameters.

|  | $\gamma$ [kN/m³] | $m$ [-] | $\nu$ [-] | $\phi$ [°] | $\psi$ [°] | $E$ [GPa] | $m_b$ [-] | $s$ [-] | $a$ [-] | $\sigma_c$ [MPa] |
|---|---|---|---|---|---|---|---|---|---|---|
| Orebody | 28.0 | - | 0.22 | - | 3 | 10 | 2.87 | 0.02 | 0.50 | 80 |
| CRF | 22.0 | 0.60 | 0.20 | 40 | 2 | - | - | - | - | - |
| Backfill | 22.0 | 0.60 | 0.20 | 40 | 2 | - | - | - | - | - |

The age-dependent parameters of the CRF were fitted against unconfined compression test results, see Figure 3. For non-cemented backfill, the same initial parameters were employed, with no age dependency. Parameters are presented in Table 2.

Table 2. Age-dependent material parameters.

| | $c_p^{ini}$ [kPa] | $c_p^{fin}$ [kPa] | $G_{ref}^{ini}$ [MPa] | $G_{ref}^{fin}$ [MPa] | $c^{ini}$ [kPa] | $c^{fin}$ [kPa] | $t^{ini}$ [day] | $t^{fin}$ [day] |
|---|---|---|---|---|---|---|---|---|
| CRF | 1 | 1000 | 35 | 700 | 10 | 750 | 10 | 56 |
| Backfill | 1 | - | 35 | - | 10 | - | - | - |

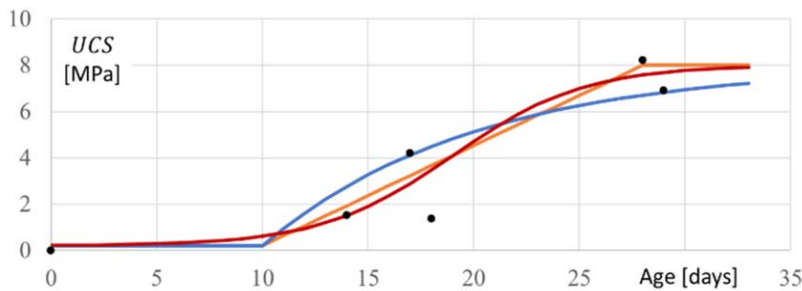

**Figure 3.** Time evolution of unconfined compressive strength. Experimental results and fits.

## 4.4. Modelling of the mining sequence

The main modelling stages can be summarized as follows: i) initial stress computed using the $K_0$ procedure; ii) sequential excavation and filling of the main stopes with CRF; iii) sequential excavation and filling of the secondary stopes with uncemented rockfill.

For the stages related to the main stopes (ii), excavation is performed by blasting and is simply modelled by deactivating the relevant cluster; filling is modelled by changing the material to cemented backfill and re-activating the cluster, at a rate of 0.66 m per day. Curing stages were added between the fill of a main stope and the excavation of the nearby secondary stopes to account for material hardening.

For the stages related to the secondary stopes (iii), the blasting and filling sequences are modelled in the same way. While the uncemented rockfill has no age-dependency, realistic time intervals were anyway used to allow the CRF to continue its hardening process when relevant. Figure 4 shows the mining sequence and the activation time of each cluster. Each time stamp is used by the model to compute ages controlling the evolution of stiffness and strength of each individual cluster with time.

| | | | | | | | | |
|---|---|---|---|---|---|---|---|---|
| | | 322 | | 238 | | 280 | | |
| | | 315 | | 231 | | 273 | | |
| | 462 | 308 | 420 | 224 | 378 | 266 | 462 | |
| | | 301 | | 217 | | 259 | | |
| | | 294 | | 210 | | 252 | | |
| | | 196 | | 112 | | 154 | | |
| | | 189 | | 105 | | 147 | | |
| | 336 | 182 | 294 | 98 | 252 | 140 | 336 | |
| | | 175 | | 91 | | 133 | | |
| | | 168 | | 84 | | 126 | | |
| | | 112 | | 28 | | 70 | | |
| | | 105 | | 21 | | 63 | | |
| | 210 | 98 | 210 | 14 | 126 | 56 | 168 | |
| | | 91 | | 7 | | 49 | | |
| | | 84 | | 0 | | 42 | | |

**Figure 4.** Construction sequence, depicted by the activation time of the clusters, in days. Gray clusters represent cemented backfill; white clusters represent uncemented rockfill.

## 4.5. Results

An example of the evolution of the functions of state variables with age is shown in Figure 5, where the shear stiffness is plotted at day 196, just after the sixth stope is backfilled. Non-uniformity of $G$ for a given cluster is explained because $G$ is a function of both age and confining pressure, which is not uniform within the cluster.

Figure 6 shows the vertical displacement $u_y$ at the end of the mining sequence, plotted at a horizontal line located just above the upper row of stopes. It is shown that the central CRF pillar undergoes settlements that are slightly larger than those of the lateral CRF pillars (colored gray), and noticeably smaller than those of the RF pillars at each side. This is an expected and realistic result, since the central pillar is the first one to be completed, and therefore the one carrying the highest load.

Figure 7 shows the distribution of vertical stress at day 460, i.e. just before the excavation and backfilling of the uppermost external stopes (see Figure 4). The effect of the mining sequence and of the different stiffness and strength of the three CRF pillars is easily noticeable. The central CRF pillar carries the highest load, the right pillar carries more load than the left pillar, and the uncemented rockfill is largely stress-free. It must be emphasized that this result was achieved without changing any material parameter throughout the modelling process.

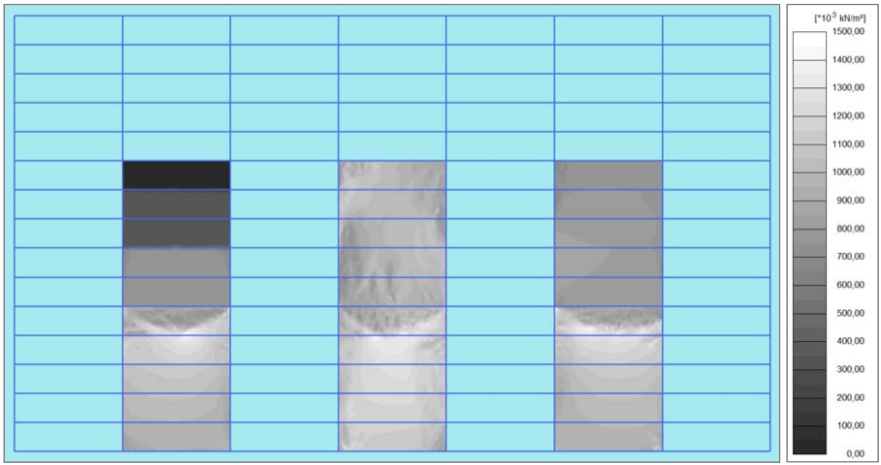

**Figure 5.** Shear modulus of the various CRF clusters at day 196, end of backfilling of stope #6. Clusters in light blue are not yet mined. Non-uniformity of G due to both age and pressure dependency.

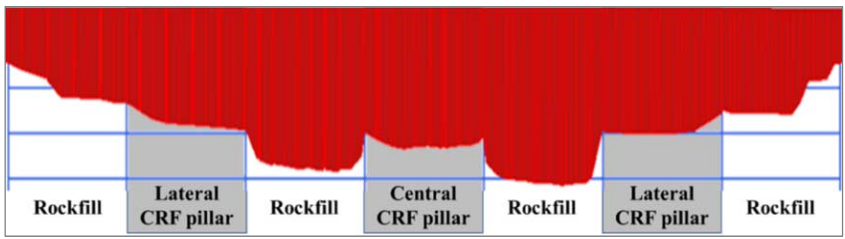

**Figure 6.** Total vertical displacements at top of orebody. Max. value 80 mm. Note the effect of the mining sequence on relative displacements of the various CRF pillars.

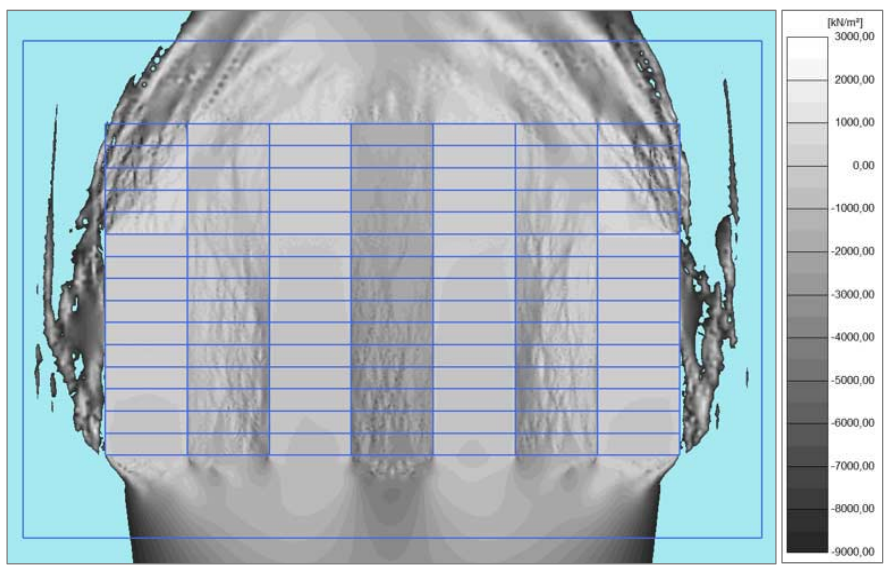

**Figure 7.** Vertical stress at day 460, just before the excavation and backfilling of the last two stopes. Note the arching effect and the higher load carried by the central CRF pillar, produced by its early construction.

## 5. Conclusions

A family of perfect plasticity constitutive models with age-, density- and pressure-dependency of strength and stiffness has been presented. Models incorporate four of the most-employed expressions for the shear modulus, five for the friction angle, and four for age-dependency reproducing cement hardening, thus producing a set of 80 options to model the behavior of cemented backfill. The family of models was implemented in Plaxis as a user-defined constitutive model; a short introduction of the algorithm was described here. Finally, a case history of the mining sequence in a South American mine was presented to prove that the model is capable of capturing the evolution of strength and stiffness of CRF pillars with age and confinement. Moreover, it is shown that a realistic simulation can be obtained while keeping material parameters of both cemented and uncemented backfills the same except for the few that account for the effect of age.

## Acknowledgements

The authors wish to acknowledge the support of the team at the University of Buenos Aires and SRK Consulting during the development, calibration and beta-testing of the models presented in this paper.